\documentclass[useAMS,usenatbib]{mn2e}

\usepackage{graphicx}

\def\iso{(age, [Fe/H])}

\title[Resolved galaxies]{Star formation histories of resolved galaxies. I. The method}

\author[E.~E.~Small et al.]{Emma E. Small$^{1}$\thanks{E-mail:
ees@astro.livjm.ac.uk}, David Bersier$^{1}$ and Maurizio Salaris$^{1}$ \\
$^{1}$Astrophysics Research Institute, Liverpool John Moores
University, Egerton Wharf, Birkenhead, CH41 1LD, UK}

\begin{document}



\maketitle


\begin{abstract}

We present a new method to determine the star formation and metal
enrichment histories of any resolved stellar system. This method
is based on the fact that any observed star in a colour-magnitude 
diagram will have a certain probability of being associated with
an isochrone characterised by an age $t$ and metallicity [Fe/H]
(i.e. to have formed at the time and
with the metallicity of that isochrone).  We formulate this as a
maximum likelihood problem that is then solved with a genetic
algorithm. We test the method with synthetic simple and complex
stellar populations. We also present tests using real data for open
and globular clusters. We are able to determine parameters for
the clusters ($t$, [Fe/H]) that agree well with results found in the
literature.  Our tests on complex stellar populations show that we can
recover the star formation history and age-metallicity relation very
accurately.  Finally, we look at the history of the Carina dwarf
galaxy using deep $BVI$ data. Our results compare well with what we
know about the history of Carina.

\end{abstract}

\begin{keywords}
Hertzsprung-Russell and colour-magnitude diagrams --
galaxies: evolution -- 
Local Group -- 
stars: evolution -- 
methods: statistical
\end{keywords}

\section{Introduction}

There are two ways to approach the complex problem of galaxy
evolution.  One approach is to look at distant galaxies using
integrated spectral energy distributions and determine broad properties
of their star formation history and metal content. This provides
a generalised view of star formation and galaxy evolution over
cosmic time. However, it is
difficult to match distant galaxies with their local counterparts. 

Another approach is to resolve individual stars in nearby galaxies.
The resulting colour-magnitude diagram (CMD) will contain information
about the age and metallicity distributions of the stars. CMD analysis
can provide a detailed picture about the star formation rate and metal
enrichment history for each individual galaxy. However, it is
difficult to draw general conclusions about galaxy evolution as this
is essentially a case by case analysis.

This second technique has seen great progress in the last two decades
and several methods have been devised to determine the star formation
history (SFH) of nearby galaxies from CMDs. Most of these methods rely
on creating synthetic CMDs to compare with the observed one using a
merit function, usually a $\chi^2$ or maximum likelihood
technique. Variants of this method have been presented by
e.g. \citet{Tolstoy96}, \citet{Harris01}, \citet{Vergely02},
\citet{Dolphin02}, \citet{Aparicio09} and \citet{deBoer12}.

Many of the methods use combinations of synthetic CMDs of ``partial''
stellar populations to reproduce different star formation
histories. The stars of both observed and synthetic CMDs are grouped
in boxes and the comparison is done on the numbers of stars in each
box.  This approach recovers both the star formation rate (SFR) and
age-metallicity relation (AMR), which together, represents the SFH of
the system. However, binning the CMD limits the resolution of the
solution and introduces a subjective element into the method by choice
of binning scheme [Aparicio \& Hidalgo (2009) note that the binning
  scheme choice can affect results].

There is still a need for an objective method that doesn't require
parametrisation or binning. \citet{Hernandez99} presented a
non-parametric method that returns the star formation rate as a
function of time, but assumes an age-metallicity relation.  In a
similar approach, \citet{Hernandez08} used a genetic algorithm to find
the best-fitting isochrone for simple stellar populations, including
distance and extinction into the fit. We approach this general problem
in a way that is similar to \citet{Hernandez08}, but extend it to
derive any type of star formation and chemical enrichment history.

We represent a composite stellar population by a linear combination of
isochrones, each with its own amplitude or weight. Therefore we are
presenting a method that can be thought of as ``multiple isochorone
fitting''. This is a full maximum likelihood analysis that directly
compares models with the observed data without the need for synthetic
CMDs, simultaneously recovers SFR, AMR, distance and extinction, and
properly weights each star in the likelihood function according to its
measurement errors.

We present the method and its implementation in
section~\ref{sec:method}. We test it on synthetic data for simple
stellar populations in section~\ref{sec:test_ssp} and on real data in
section~\ref{sec:clusters}, where we look at several open and globular
clusters.  Section~\ref{sec:test_csp} shows how we can recover complex
star formation histories and section~\ref{sec:carina} shows a
comparison of our method with results obtained by other groups for the
Carina dwarf spheroidal galaxy.  We present a summary and conclusions
in section~\ref{sec:conlusions}.

\section{Maximum Likelihood Method}
\label{sec:method}

\subsection{Likelihood Function}

As discussed above, the stellar population content of any stellar
system can be modeled as a linear combination of isochrones, each
being characterised by its age and metallicity (t,
[Fe/H]])\footnote{Helium content also affects stellar properties but
    all authors assume that $Y$ (helium content) scales linearly with
    metallicity Z.  In this paper we use $\Delta Y/\Delta Z = 1.4$
    from \citet{Pietrinferni04}. We also use the observable [Fe/H]
    rather than Z for the metal abundance,; see \citet{Pietrinferni04}
    for the conversion.}.
When reconstructing the SFH of the system, the specific problem is to
use the information in the CMD in order to access the SFR as a
function of time. Our approach is to calculate the relative
probability that each star ``originated'' from a particular isochrone.
This probability can be defined by the distance in magnitude space
between an isochrone and the actual position of the star, accounting
for the measurement errors.  Formally, the problem then is to find the
weight of each isochrone in the star formation history, i.e.  the
relative number of stars that each isochrone contributes to the
CMD. We will build a likelihood function where the free parameters
(i.e. the parameters that maximise the likelihood) are these isochrone
weights.

We denote the magnitudes of a star $j$ in two passbands as $A_j$ and
$B_j$. Corresponding theoretical magnitudes from isochrone $i$ will be
denoted $A_{iM}$ and $B_{iM}$ respectively. These $A_{iM}$ and $B_{iM}$ are
shifted to account for the distance modulus and extinction of the
observed system. The quantity
\begin{equation}
E(A_{ij}) = \frac{1}{2\pi\sigma_A}
\exp\left[-\frac{(A_j-A_{iM})^2}{2\sigma_A^2}\right]
\label{eq:errors}
\end{equation}
represents the distance in magnitude $A$ between the observed star $j$
and a single point on isochrone $i$, assuming a gaussian error
$\sigma_A$ on $A_j$. $\sigma_A$ is given by
\begin{equation}
\sigma_A = \sqrt{\sigma_{A,\mathrm{phot}}^2+\sigma_{A,\mathrm{iso}}^2}
\label{eq:sigma}
\end{equation}
where $\sigma_{A,\mathrm{phot}}$ is the photometric error and
$\sigma_{A,\mathrm{iso}}$ accounts for the differences in magnitude
between two neighbouring isochrones. See sec~\ref{sec:implement} for
the rationale behind the use of $\sigma_\mathrm{iso}$.

Since the mass of star $j$ is unknown, the probability function has to
be an integral over all possible masses along each isochrone. The
probability will also need to be weighted by the relative number of
stars along the isochrone, i.e. the IMF $f(M)$, and a function
describing the observed photometric completeness $c(A,B)$.  The
relative probability $p_{ij}$ that a given star $j$ ``belongs'' to
an isochrone $i$ is
\begin{equation} 
p_{ij} = \frac{1}{C_i} \int_{M_{il}}^{M_{iu}}
E(A_{ij}) E(B_{ij}) c(A,B) f(M)dM
\label{eq:prob}
\end{equation}
where $M_{iu}$ and $M_{il}$ are the upper and lower mass limits on the
isochrone $i$ and
\begin{equation}
C_i = \int_{M_{il}}^{M_{iu}} c(A,B) f(M)dM.
\end{equation}
is a normalisation factor for the completeness and initial mass
function. 

In the above definition of the probability function we use two
magnitudes, $A$ and $B$. However this can easily be adapted to include 3
or more magnitudes. Since we consider the measurement of each
magnitude to be independent, we can just include another term
$E(C_{ij})$ describing magnitude $C$ and adjust the completeness
function accordingly. The probability function for 3 magnitudes
becomes:
\begin{equation} 
p_{ij} = \frac{1}{C_i} \int_{M_{il}}^{M_{iu}}
E(A_{ij}) E(B_{ij}) E(C_{ij}) c(A,B,C) f(M)dM
\label{eq:prob_3mag}
\end{equation}
This will allow all photometric information for the data available for
each star to be used to constrain the star formation history.

The next step in constructing our likelihood function 
is to determine the relative probability that a single star $j$
belongs to a complex stellar population, which is represented by a
linear combination of SSPs with different weights. The combined
probability $p_j$ is given by
\begin{equation}
p_j = \sum_{i=1}^{n_i}a_ip_{ij}
\end{equation}
where $a_i$ represents the weight of each isochrone and $n_i$ is
the total number of isochrones in the library.  The weights are the
relative number of stars each isochrone contributes to the CMD and are
subject to the following constraints:
\begin{equation}
  \sum_{i=1}^{n_i}a_i = 1
\end{equation} 
and
\begin{equation}
  0 \leq a_i \leq 1.
\label{eqn:lim}
\end{equation} 

The likelihood function is defined as the product of probabilities
of all stars in the CMD:
\begin{equation}
L = \prod_{j=1}^{n_j}p_j = \prod_{j=1}^{n_j} \left[ \sum_{i=1}^{n_i} a_ip_{ij} \right]
\end{equation}
\begin{equation}
\ln\left(L\right) =  \sum_{j=1}^{n_j}\ln\left(\sum_{i=1}^{n_i}a_ip_{ij}\right) .
\end{equation}

One can then use a maximisation technique to determine the weights of
each isochrone, and hence the star formation and metal enrichment
histories.  Our method maximises the log-likelihood using a genetic
algorithm to find the combination of weights $a=(a_1, a_2,\ldots,
a_{n_i})$, that most likely produced the observed CMD.

In order to interpret this as a star formation history, the weights for
each isochrone need to be rescaled to account for those stars that are not in 
the CMD due to incompleteness effects, and those stars that have evolved beyond 
the final stage covered by the isochrones (either the end of the asymptotic giant branch 
phase or central carbon ignition, 
depending on the age). 
For instance, if a ``galaxy'' has had two bursts of star
formation of same intensity (i.e. same total mass formed in each), one
12 Gyr ago and the other 100 Myr ago, there will be many more stars in
the CMD from the young component. All stars with $M\ga 0.8 M_\odot$
from the old component will have disappeared (they have evolved beyond the end 
of the asymptotic giant branch) and the young burst will
appear much stronger than the old one. 
The number of missing stars in the CMD for
each isochrone can be estimated from the IMF and completeness
functions and added to its weight. The weights are then renormalised
to give the correct relative number of stars formed at each age and
metallicity.

\subsection{Genetic Algorithm}

The maximum likelihood solution is found using a genetic algorithm.
This is a global optimization technique, inspired by the concept of
evolution. Here we will just give a brief outline of the general
procedure.  To begin with, a population of randomly generated
potential solutions are created and the fitness of each member is
evaluated. Each individual solution is composed of a sequence of
parameters/genes, which is referred to as its genotype.  Members from
the current generation are selected, with a probability proportional
with their fitness ranking, to 'breed' and produce the next generation
of solutions. The breeding process involves two operations: crossover
and mutation. The crossover operator randomly combines the genes of
the two parents, allowing for the passage of information between
generations. The mutation operator randomly replaces existing genes
with new genes, allowing for large leaps in the parameter space and
prevents the algorithm being trapped by local maxima.  This process is
iterated over many generations until the algorithm converges to the
global maximum.

For this problem, the genotype of each solution is the list of weights
for the isochrones that characterise the star formation history.
The fitness is evaluated by the likelihood. Since the parameters are
linearly constrained by equation \ref{eqn:lim}, our genetic algorithm
adopts the same approach as \cite{Michalewicz91}. The weights
are represented as an array of floating point numbers and the
crossover and mutation operators are modified to preserve the linear
constraints between generations. We have found this approach to give a
faster and more reliable convergence to the global maximum than a
traditional genetic algorithm as found in \citet{Charbonneau95}.

\subsection{Error Analysis}

Uncertainty in the star formation history will arise from stars in the
population being associated with the wrong isochrone due to overlap of
isochrones in the CMD, imperfect models, observational effects and
uncertainty in distance and reddening values. The fact
that we are using a discrete set of isochrones to model a continuous
range of age and metallicity will also constribute to the
uncertainty.Thus we need to calculate confidence intervals 
for the weights of each
isochrone.

Isochrone weights with similar age and metallicity will be very
highly correlated in the solution (because they are close in the 
CMD). They, therefore, cannot be treated as independent parameters.
For this reason we estimate the confidence intervals of
all the weights simultaneously using a Monte Carlo method.  In a large
sample limit, the likelihood function approximates an $n$-dimensional
Gaussian at the global maximum, where $n$ is the number of parameters
(i.e. number of isochrones).  Following this, it can be assumed that a
$n$-dimensional confidence region $Q$ will have a chi square
distribution. We can determine $Q_\gamma$, as the limit of confidence
region, which has a coverage probability $(1-\gamma) = 0.683$ and thus
define a lower limit to the log-likelihood using
\begin{equation}
\ln L_{\mathrm{lim}} = \ln L_{\mathrm{max}} - \frac{Q_\gamma}{2} .
\end{equation} 
The limit in log-likelihood corresponds to solutions of the SFH
$1\sigma$ from the maximum. We use a Monte Carlo approach to generate
solutions with a likelihood above this limit and determine the range
in weights for each isochrone.

The limit in log-likelihood $\ln L_{\mathrm{lim}}$ is determined by
the number of free parameters.
Although our SFH is initially evaluated using a full library of
isochrones, covering the largest possible range of age and metallicity,
typically only a subset will actually contribute to the maximum
likelihood solution. We will only include isochrones that have a
weight greater than $0.1\%$ in our error analysis. Isochrones with a
non-zero weight below this limit do not contribute significantly to
the SFH solution and are most likely just a statistical artefact of
the method. This statistical artefact is seen for data with small
photometric errors, where individual stars that are associated with
low-weight isochrones have more of an impact on the likelihood 
function. The inclusion of isochrones with weights $\leq 0.1$\%  
in the error analysis has 
the effect of increasing the errors on all isochrones in the solution
by lowering $\ln L_{\mathrm{lim}}$ rather than being a genuine source
of variance for the distribution of stars in themselves.

Generally speaking, the greater the photometric error, the greater 
the confidence limits. This is due to the fact that increasing the
photometric error decreases the ability of the probability to
distinguish between neighbouring isochrones. Although it is essential
to evaluate the star formation history using all the isochrones to
achieve the highest resolution in age and metallicity possible, the
resolution may be too high, depending on the quality of the data, to
effectively and accurately present the resulting star formation
history.  This is particularly an issue for old stars, where there is
very little difference between neighbouring isochrones.  One way of
reducing errors in the weights and therefore uncertainty in the star
formation history solution is to combine isochrones into age and
metallicity bins, where the weight of each bin is the sum of the
weights of the isochrones it contains.
Error bars on individual weights (before binning) may be large because
stars can move between neighbouring isochrones; when we bin
isochrones however, stars can move inside a single bin but it's more
difficult to move outside of it.

\subsection{Implementing the Method}
\label{sec:implement}

In the implementation of this method we use BaSTI isochrones
\citep{Pietrinferni04}.  These densely sampled isochrones each consist
of 2000 mass points, with a fixed number of points for each
evolutionary phase (e.g. MS, SGB, RGB, HB).  It is important to note
that the probability function requires an integration along the
isochrone, which is estimated using a discrete set of points rather
than a continous line. This means that for CMD data with small
photometric error, the sampling of the isochrones can become
significant in determining the probabilities. This is an important
reason why we chose this particular grid of isochrones. Poor sampling
can actually result in stars not registering a significant probability
for the correct isochrone because they lie midway between two points.
This is particularly a problem for the Main Sequence of young
isochrones where a relatively small number of mass points cover a
large extent in magnitude space. Since the MS is the area of greatest
overlap between isochrones with the same metallicity but different
ages, insufficient sampling of points on young isochrones can bias the
recovered star formation history to higher ages.  We find that the
magnitude spacing between adjacent points on the isochrone should be
no greater than the photometric error $\sigma$ of the star, to give a
reasonable approximation of the probability. We increase the sampling
of points through linear interpolation so that the spacing is less
than $\sigma/2$ when integrating over the isochrone.

The isochrone grid we use has 51 ages ranging from $30$ Myrs to $13$
Gyrs and 11 metallicities ranging from [Fe/H]~$=-2.27$ to
[Fe/H]~$=0.4$. As discussed previously, our method is attempting to
map a continuous distribution of possible ages and metallicities on a
discrete grid of isochrones. This simplification necessitates an
additional error term in our probability function, which we have
denoted $\sigma_\mathrm{iso}$. Specifically $\sigma_\mathrm{iso}$ is
the minimum error required for each star in the CMD to register a
non-zero probability with the correct (i.e. nearest) isochrone in the
grid.  Ideally $\sigma_\mathrm{iso}$ would also incorporate the
uncertainty in photometric magnitudes due to errors in the models as
well as that due to the quantisation in age and metallicity, however
this is very difficult to quantify. For now, we take a simple estimate
of $\sigma_\mathrm{iso}$ based on our isochrone grid. We use
$\sigma_\mathrm{B,iso}=0.015$, $\sigma_\mathrm{V,iso}=0.03$ and
$\sigma_\mathrm{I,iso}=0.019$, which is the average difference in
magnitude of a MS star with the same age and mass but neighbouring
metallicities. We only estimate these ``errors'' using differences in
metallicity because this has a much greater effect on the magnitude of
stars than age, given the isochrone grid we use.  The inclusion of
$\sigma_\mathrm{iso}$ is very important for datasets with small
photometric error (see sec~\ref{sec:m37} and sec~\ref{sec:carina}). It
ensures that stars with small photometric uncertainty do not ``fall
through the cracks'' in our isochrone grid and that the resulting
maximum likelihood solution will be a product of all the stars in the
CMD.

Another point to note about the method is that the fit will be
dominated by the lowest mass stars.  This is due to the IMF term in
the probability function and the fact that these stars have hardly
evolved at all.  For sufficiently deep data these stars will reside
in the area of greatest overlap between isochrones of different ages
in the CMD. Therefore the lower MS will contain relatively little
information distinguishing age in complex stellar populations. (The
effects of metallicity on the MS can be mimicked by reddening). We
apply a magnitude cut to the isochrones and the CMD to remove the
lower MS stars. We ensure that this cut is taken below the MS turnoff
of the oldest stars in the population.  This increases the influence
of the more discriminating post MS evolutionary stages in determining
the star formation history.

\subsection{Observational effects}
\label{sec:biases}

There are two effects that we need to consider before going further.
In most real situations, crowding is sufficiently important that one
needs to perform artificial star experiments.  Such tests consist in
``injecting'' stars in CCD images at random locations, doing
photometry on the images, and comparing the input magnitudes with the
recovered magnitudes. While these tests are usually done to determine
the completeness level of the data, they also reveal the extent of the
biases that affect the magnitudes and the associated errors. \\
\textit{Photometric errors:}
In crowded fields, photometry is done by fitting a point-spread
function to the stellar images. The uncertainties returned by a
photometry program do not account for the fact that nearby stars
contaminate the fit. Numerous artificial star experiments have shown
that the true errors are significantly larger than what is returned by
a photometry program \citep{AG95, Olsen03} and, at any given
magnitude, they are not necessarily Gaussian in shape. Our formulation
(eq.~\ref{eq:errors}) uses a Gaussian function but it does not need to
be so. In a real case, we would perform artificial star tests and
determine the magnitude dependence of the error law and its shape
(i.e. accounting for the fact that, at a given magnitude, errors are
not symmetrically distributed). Knowing this, we can modify the code
to account for the known distribution of errors. \\
\textit{Magnitude bias:}
Artificial star tests also reveal a magnitude bias in the sense that
the recovered magnitudes are brighter than the input magnitudes
\citep{AG95, Olsen03, Stephens01}. Furthermore, the deviation gets
larger and larger when the magnitude gets fainter and fainter.  In our
method, we would not correct the observed magnitudes but we would
change the input isochrones.  Knowing the extent of the bias as a
function of magnitude, we can correct the magnitudes in each
isochrone, i.e. subtract the appropriate deviation as a function of
magnitude.

Together, these two effects would have to be implemented differently
for each particular stellar system we consider. We have not done this
in this paper because we focused on the method itself and its
accuracy. The exact shape of the error law (i.e. what goes in
eq.~\ref{eq:errors}) and the exact shapes of the isochrones are not
relevant for the tests we describe in this paper.

\section{Simple Stellar Populations}
\label{sec:test_ssp}

In this section and section~\ref{sec:clusters}, we perform tests to
confirm the validity of the method for the special case of simple
stellar populations. Simple stellar populations are the product of a
single burst of star formation, where all the stars can be considered
to have the same age and metallicity. In other words, all the stars in
the population originate from a single isochrone.  The problem of
determining the star formation history simplifies to using the
likelihood function to identify the best-fitting isochrone. This
version of the method is very similar to a method used to determine
the parameters of star clusters devised by
\citet{Hernandez08}. However, unlike these authors we are not
attempting to determine these parameters precisely, but to identify the
closest age, metallicity, distance and reddening from a preset grid of
isochrones. For this reason and the fact that ranges of distance and
reddening values can usually be constrained from the literature, it is
not necessary to include optimisation from a genetic algorithm. We
simply calculate the likelihood of all the possibilities in our
reduced parameter space.

We test the method using both synthetic data in this section and on
cluster data in section 4. The advantage of using synthetic data is
that the underlying star formation history is known. Also since the
CMD is created and analysed using the same models, we can properly
test the strengths and weaknesses of the method without having to
factor in issues with the stellar evolutionary models.  We generate
our synthetic CMDs using the BaSTI webtools \citep{Cordier07} and,
unless otherwise stated, the solar-scaled no-overshooting set of
isochrones. For these tests we use $V$ and $I$ magnitudes.

\begin{figure}
\centering
\includegraphics[scale=0.4]{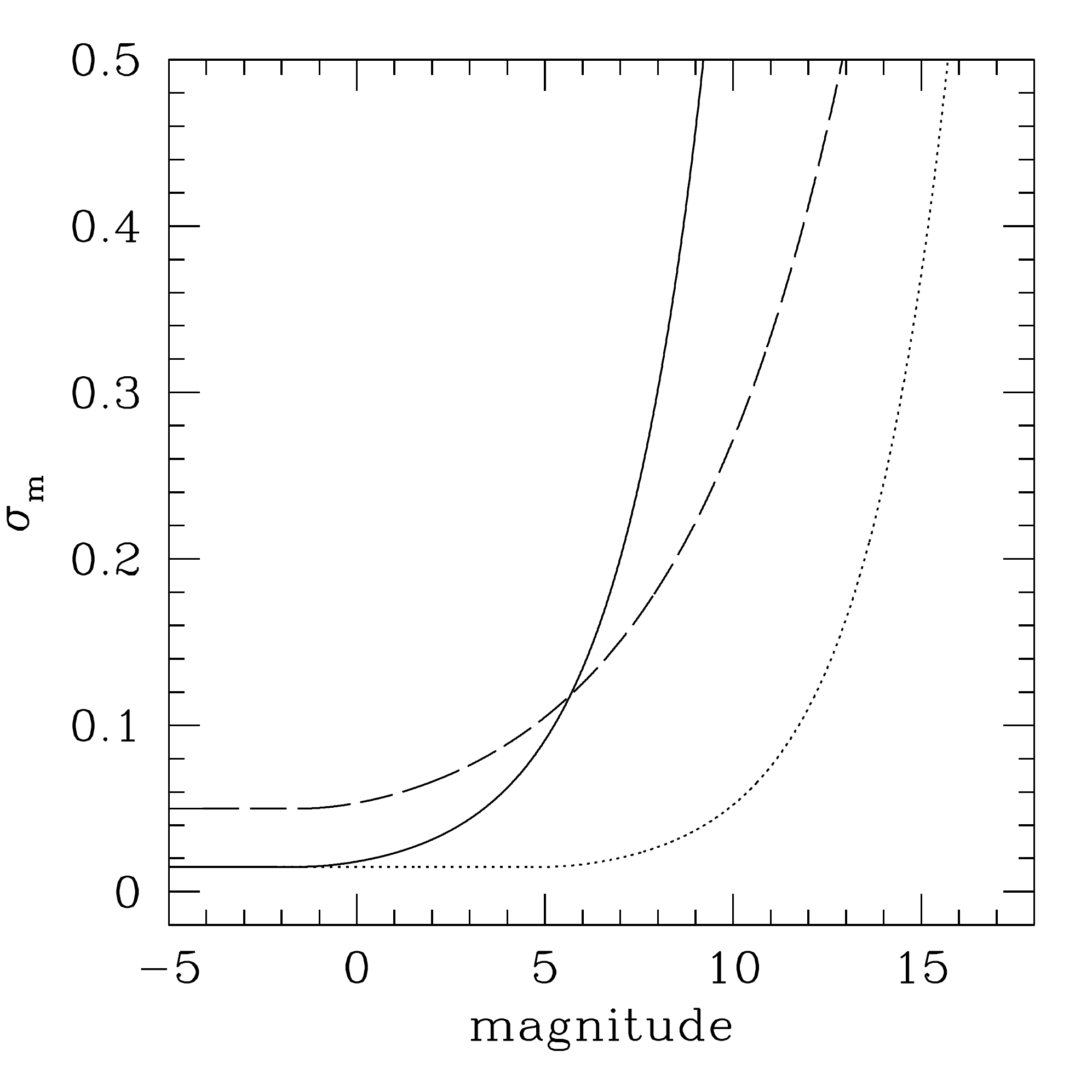}
\caption{Effects of various parameters on the adopted error law. The
  solid line is eq.~\ref{eq:error_law}, the dashed line is 
  eq.~\ref{eq:error_law} but with $\beta=0.27$ and $\sigma_A^0 =
  0.05$, the dotted line is  eq.~\ref{eq:error_law} but with
  $A_0=5$.}
\label{fig:errorlaw}
\end{figure}

Photometric errors on magnitude $A$ were simulated using the following
typical error law:
\begin{eqnarray}
\sigma_A & = & \sigma_A^0  \hspace*{40mm} \mathrm{if} A\leq A_0 \nonumber \\
         & = & \sigma_A^0\, \frac{e^{\beta(A-A_0)}}{1+\beta(A-A_0)}
               \hspace*{20mm} \mathrm{if} A\geq A_0 
\label{eq:error_law}
\end{eqnarray}
where $\beta = 0.5$, $\sigma_A^0 = 0.015$ and unless otherwise stated
$A_0=-1.5$. Examples of this error law are shown in
Fig.~\ref{fig:errorlaw}.

We use synthetic simple stellar population data to test that the
method can recover the correct isochrone for an old (12 Gyr, -1.27)
intermediate age (2 Gyr, -1.27) and young (100 Myr, -0.25) SSP. As a
self-consistency check, we created a CMD for each of these SSP (with
photometric errors) and verified that the method recovered the input
parameters.  We then perform tests to investigate how the HB,
binaries, blue stragglers, the IMF and completeness in the CMD affects
the results.  In these cases the distance and reddening are assumed to
be known and are not solved for simultaneously.

\subsection{HB Morphology}

\begin{figure}
\centering
\includegraphics[width=75mm, scale=0.9]{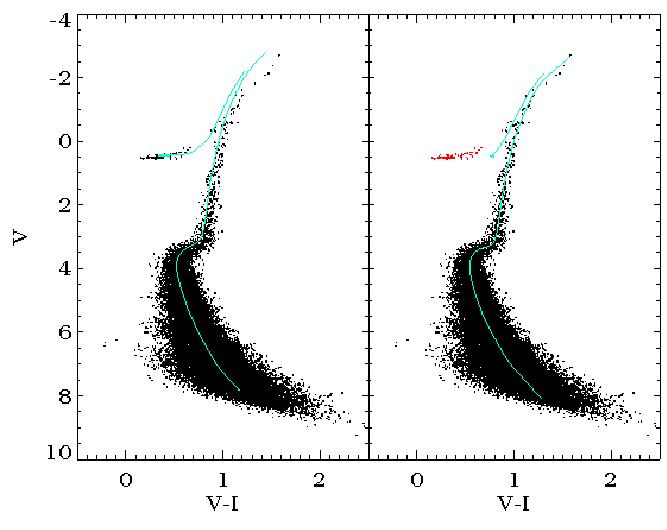}
\caption{HB test: CMD of SSP (input t=11 Gyr, [Fe/H]~$=-1.27$) with
  output isochrone for tests \textit{i} and \textit{ii} (t=13 Gyr,
  [Fe/H]~$=-1.79$) on the left and tests \textit{iii} and \textit{iv}
  (t=11 Gyr, [Fe/H]~$=-1.27$) on the right. The red points in
the right panel are those stars that were removed for tests \textit{iii}
and \textit{iv}}
\label{fig:hb}
\end{figure}

\begin{table}
\caption{Results for HB tests. The input isochrone has (11 Gyr, $-$1.27)}
\begin{tabular}{l c l}
\hline
test &  t (Gyrs)  & [Fe/H] \\
\hline
\textit{i} & 13 &   $-1.79$\\
\textit{ii} & 13 &  $-1.79$\\
\textit{iii} & 11 & $-1.27$\\
\textit{iv} & 11 &  $-1.27$\\
\hline
\end{tabular} 
\label{tab:hb}
\end{table} 

The location of Helium burning stars in the CMD is not only dependent
on the age and initial chemical composition of the population, but
also on the mass loss on the RGB.  As a consequence, old stellar
populations of the same age can display different HB morphologies. In
particular, mass loss is poorly understood and its treatment is a
shortcoming of stellar models. The BaSTI isochrones use an empirically
derived fixed law with a free parameter $\eta$ either set to $0.2$
or $0.4$ \citep{Pietrinferni04} to simulate mass loss. To test the
robustness of the method against data with a different HB morphology,
we create a CMD with $\eta=0.4$ and solve for isochrones with
$\eta=0.2$. In this case the isochrones are underestimating the amount
of mass loss occuring on the RGB and subsequently the HB phase on the
isochrone is redder than the CMD data. For this test, the synthetic
SSP was generated with an age of 11 Gyrs and metallicity [Fe/H]~$=-1.27$.
We evaluate the parameters of the
SSP with \textit{i)} the full isochrones and the CMD, \textit{ii)}
the HB phase removed from the isochrones, \textit{iii)}  HB stars
removed from CMD but using the full isochrones, and \textit{iv)}
removing HB phase from isochrones and HB stars from CMD.  The results
are presented in Table~\ref{tab:hb} and Fig.~\ref{fig:hb}.

The results show that HB morphology can cause the method to
select the wrong isochrone. In this case, a bluer HB leads to 
a lower metallicity and greater age. 
However it appears that the method can perform the fit successfully
only using MS, SGB and RGB phases since the correct isochrone was
selected when the HB stars were removed from the dataset.

\subsection{Unresolved Binaries}
 
In the CMD, unresolved binaries are most evident on the MS, where they
form a second, brighter and redder sequence of stars.  For post MS
phases, the light of the giant star will completely dominate that of
its MS companion. In order to test the effects of unresolved binaries
we generated CMDs with different binary fractions, from 0.1 to 0.5 for
our young, intermediate and old SSPs. The mass ratio follows a flat
probability distribution between 0.7 and 1. The results for 0.2 and
0.5 are summarised in Table~\ref{tab:bin}.

\begin{figure}
\centering
\includegraphics[width=90mm, scale=0.9]{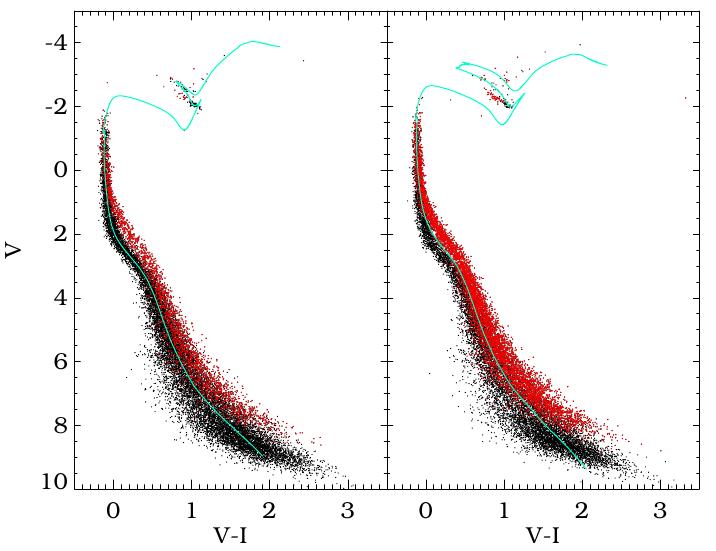}
\caption{CMD of SSP (input:t=100 Myr, [Fe/H]~$=-0.25$) with selected isochrones
for binary fraction =0.2 (output: t=100 Myr, [Fe/H]~$=-0.25$) on the left and 0.5 
(output: t=80 Myr, [Fe/H]~$=$0.06) on the right. The binary stars are indicated
by red points.}
\label{fig:bin}
\end{figure}

\begin{table}

\caption{Results for different unresolved binary fractions}
\begin{tabular}{l l l l l r}
\hline
\multicolumn{3}{c}{Input} &  & \multicolumn{2}{c}{Output} \\
\hline
t (Gyrs)  &  [Fe/H] & binary &  & t (Gyrs) & [Fe/H] \\
  &   & fraction &  &  \\
\hline
0.1 & $-0.25$ & 0.2 & & 0.1  & $-0.25$ \\
0.1 & $-0.25$ & 0.5 & & 0.08 & $ 0.06$ \\
2   & $-0.25$ & 0.2 & & 1.75 & $-0.25$ \\
2   & $-0.25$ & 0.5 & & 1.75 & $-0.25$  \\
12  & $-1.27$ & 0.2 & & 12   & $-1.27$ \\
12  & $-1.27$ & 0.5 & & 11.5 & $-1.27$ \\
\hline
\end{tabular}
\label{tab:bin}
\end{table}

The method appears to be relatively robust against binaries for the
intermediate and old SSPs, but not for the young population.  For a
high binary fraction, the method selects an isochrone with a higher
metallicity and lower age than the input values (see
Fig.~\ref{fig:bin}). This is unsuprising since the post-MS
evolutionary phases are not well populated for a young SSP and the
results are more dependent on fitting the MS. However, if we take a
relatively high MS cut (e.g. just below the MSTO), then the post
evolutionary phases have more weight in the fit and the method selects
the correct isochrone.

\subsection{Blue Stragglers}

Blue stragglers are bright blue stars that inhabit the region just
above the Main Sequence Turn Off in a CMD. There are two possible 
mechanisms suggested for blue straggler formation: stellar 
mergers by collision \citep{Bailyn95} and mass transfer and 
coalescence in binary systems \citep{Carney01}. In galactic
globular clusters, the frequency of blue stragglers normalised 
to the number of HB stars varies between 0.1 and 1 \citep{Piotto04}.

It is necessary to test the robustness with respect to the presence of
blue stragglers in the data set because they lie close to the MSTO,
which is the main age indicator in terms of isochrone fitting. We
generate an SSP with \iso = (12~Gyrs, $-$2.27) and add the maximum
fraction of blue stragglers (i.e. a ratio of BS to HB stars of 1).  We
create our blue stragglers by taking stars from the MS of an SSP with
\iso = (5~Gyrs, $-$2.27) with a range in $V$ from the MSTO magnitude
to 1 magnitude above the MSTO in our original CMD. The method selected
the correct isochrone, suggesting that blue stragglers are not
numerous enough to affect the fit of the MSTO.

\subsection{IMF and Completeness}

The IMF and completeness both affect the distribution of stars along
different evolutionary phases in the CMD. In the method the IMF is
assumed to be a power law with an exponent of $-2.35$. To check that
the method is robust against an input IMF we generate an SSP with a
power law IMF and solve with exponents of $-1.35$, $-2.35$, $-3.35$. We
find that the correct isochrone was selected each time, although the
value of the maximum likelihood was highest for the correct IMF. 

Completeness on the other hand is a function of magnitude and can be
different among different isochrones. We simulate the effects of
completeness in the CMD using the following function:
\begin{eqnarray}
c(A)=\frac{1}{1+\exp(A-A_c)/\Delta A}
\label{eq:comp}
\end{eqnarray}
for values of $A_c$ between 6 and 1  
and $\Delta A =$~0.1 and 0.5.
We also shift $A_0$
in the error law to correspond to a photometric error $\sigma_A \sim
0.3$ for the lowest magnitudes present in the CMD for a more realistic
simulation of photometric completeness. 
Our tests  show that, in the case of SSPs, the method is robust against an
incorrect completeness function and that the method will recover the
correct isochrone for any magnitude cut taken below the MSTO.

\section{Clusters}
\label{sec:clusters}

We now test the method on real star clusters, to ensure that the method
can recover reasonable values with real data and that we can compare
our results with other methods.  We have chosen 3 clusters that are
varied in age and metallicity; an intermediate age open cluster, M~37;
a globular cluster, M~3; and an old, metal-rich open cluster, NGC~6791.
We allow the distance modulus and extinction of the population to
vary, with ranges of values constrained from the literature.  In this
case we want the method to select the nearest isochrone, distance
modulus $\mu_0 = (m-M)_0$ and reddening $E(B-V)$ to the actual
parameters of the cluster.

\subsection{M~37}
\label{sec:m37}

M~37 is a well studied intermediate age open cluster, with parameters
determined by several groups summarised in \citet{Hartman08}. We use
$BV$ data\footnote{downloaded from WEBDA at www.univie.ac.at} from
\citet{Kalirai01}, with cluster members selected by
\citet{Hartman08}. The final CMD consists of 1473 stars. We note that
the photometric error for the M~37 data are very small, therefore the
systematic error dominates the error we put into the likelihood
function.  Since the CMD for M~37 contains very few post MS stars and
is very deep, we fit isochrones using different magnitude cuts from
$m_V=21$ to $m_V=14$. The completeness at $m_V = 21$ is estimated to
be 0.94 \citep{Kalirai01}, therefore we use a simple step function at
our lower $V$ magnitude cutoff as a completeness function.  We also
use models that include overshooting, since these are more appropriate
models for clusters with ages in the range of M~37.  Our initial
ranges for $\mu_0$ and $E(B-V)$ are based on the literature, with
$\mu_0 = \{10,10.1,\ldots,11\}$ and $E(B-V) = \{0.2,0.21,\ldots,0.3\}$.

\begin{table}
\caption{Results for M~37}
\begin{tabular}{l c r c c}
\hline 
   &    t (Myrs)  & [Fe/H] & $\mu_0$ & $E(B-V)$ \\
\hline
$m_V=21-20$ & 500  & $0.06$ & 10.9 & 0.21 \\
$m_V=19-14$ & 500  & $-0.25$ & 10.6 & 0.29 \\
\hline
\end{tabular}
\label{tab:M37}
\end{table}

\begin{figure}
\centering
\includegraphics[scale=0.6]{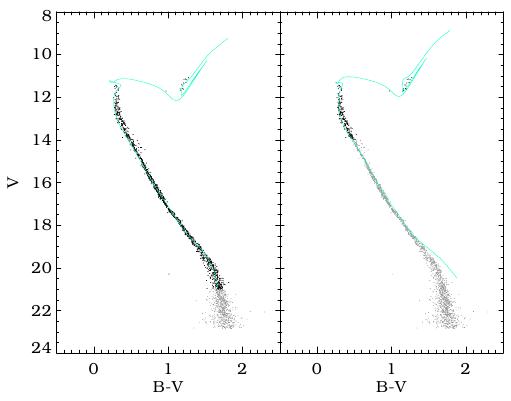}
\caption{CMD of M~37 with maximum likelihood isochrone when $m_V=14$ (left) and
$m_V=21$(right), grey points indicate stars that have been removed before
likelihood analysis}
\label{fig:m37_cut}
\end{figure}

The results for M~37 are summarised in Table~\ref{tab:M37} and plotted
in Fig.~\ref{fig:m37_cut}. 

Different groups have obtained slightly different values for the
parameters of this cluster \citep{Hartman08,Kalirai01,Sills00}.  The
values we recover for the cluster parameters are in broad agreement
with those found in those papers. A self consistent test here is 
to compare the results with the values deternined by 
Salaris et al. (2009) because they are based on the same
set of isochrones but used a different method.  The distance was
determined empirically from MS fitting and the age came from fitting
TO absolute magnitudes with BaSTI isochrones, assuming
[Fe/H]$\ =-0.20$ and $E(B-V)=0.23$ from \citep{Kalirai05}.  Our values
of age, metallicity and apparent distance modulus are consistent with
those of \citet{Salaris09}.

\subsection{M~3}

\begin{figure}
\centering
\includegraphics[scale=0.5]{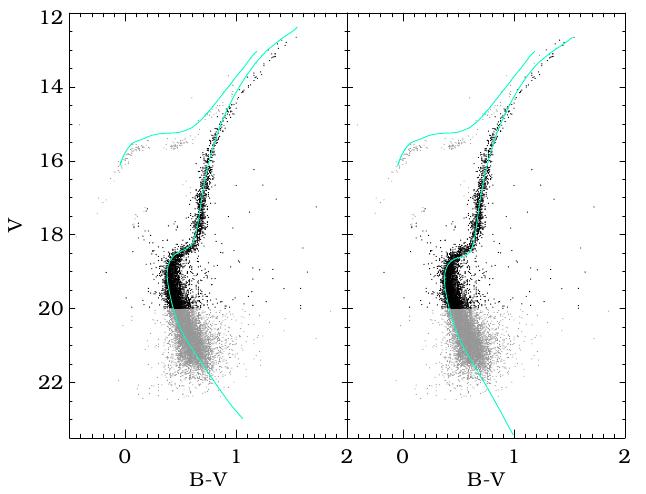}
\caption{CMD of M~3 with maximum likelihood solar-scaled isochrone (left)
  and maximum likelihood $\alpha$-enhanced isochrone (right). Stars in
  grey have not been used in the fit.}
\label{fig:m3}
\end{figure}

The data for M~3 are taken from \citet{Buonanno94}, with
photometric errors estimated from Fig.~7 in that paper. The CMD is
displayed in Fig.~\ref{fig:m3}, where grey points represent stars that
have been removed from the CMD prior to analysis. We have estimated
the completeness function using Table~7 in \citet{Buonanno94} and
have also taken a lower MS magnitude cut at $m_V=20.05$. The CMD features
an extended Horizontal Branch, which has been removed and we only use
points up to the tip of the RGB in the isochrones. The final CMD
consists of 4173 stars.

We complete the test using both solar scaled and $\alpha$-enhanced
isochrones. The ranges in distance modulus and extinction are 
$\mu_0 = \{14.75,14.8,\ldots,15.3\}$ and $E(B-V)=\{0,0.01,\ldots,0.07\}$. 
The results are presented in Table~\ref{tab:M3} and Fig.~\ref{fig:m3}.

\begin{table}
\centering
\caption{Results for M~3}
\begin{tabular}{l c c c c}
\hline 
  &    t (Gyrs) & [Fe/H] & $\mu_0$ & E(B-V) \\
\hline
 ss & 13.0 & $-1.49$ & 15.05 & 0.0 \\     
 ae & 10.5 & $-1.31$ & 15.3 & 0.0 \\     
\hline
\end{tabular}
\label{tab:M3}
\end{table}

Visual inspection of Fig.~\ref{fig:m3} show that the method provides a
reasonable fit of isochrone to the data for both solar scaled and
$\alpha$-enhanced models, although the upper RGB is not well
represented by the isochrones in both cases. The solutions for the
$\alpha$-enhanced models have a likelihood greater than the
scaled solar ones, suggesting a better fit,consistent with 
expectation, for spectroscopy of M~3 stars shows they are
$\alpha$-enhanced with [$\alpha$/Fe]~$=0.27\pm 0.03$
\citep{Carney96}.

The result from the $\alpha$-enhanced isochrones is consistent with 
literature values of the cluster parameters, e.g. 
the recent values found by \citet{Benedict11}, who determine
$\mu_0 = 15.17\pm 0.12$ from an empirical calibration of the RR Lyrae
distance scale (using $E(B-V)=0.01$ and [Fe/H]~$=-1.57$)
and an age of 10.8~Gyr \citet{Benedict11}.

\subsection{NGC6791}

The data for NGC6791 are taken from \citet{Stetson03}. Since there are
measurements in $B$, $V$ and $I$, we take this opportunity to
determine and compare the results using different magnitude
combinations. We solve for $BV$, $VI$ and $BVI$. There is evidently
foreground contamination in the CMD for NGC6791, however we do not
attempt to remove this in order to test whether the method is robust
against contaminating stars.  We take a magnitude cut at $m_V=20$, the
resulting input CMDs consist of 12763 stars.  The ranges in distance
modulus and extinction are $\mu_0 = \{12.5,12.6,\ldots,13.5\}$ and
$E(B-V)=\{0.1,0.12,\ldots,0.2\}$.

\begin{table}

\caption{Results for NGC6791}
\begin{tabular}{l c c c c}
\hline 
   &    age (Gyrs)  & [Fe/H] & $\mu_0$ & $E(B-V)$ \\
\hline
$BV$  & 8.5 & 0.40 & 13.0 & 0.14 \\   
$VI$  & 9.0 & 0.26 & 13.1 & 0.12 \\   
$BVI$ & 8.5 & 0.40 & 13.1 & 0.12 \\   
\hline
\end{tabular}
\label{tab:ngc6791}
\end{table}

\begin{figure}
\centering
\includegraphics[scale=0.6]{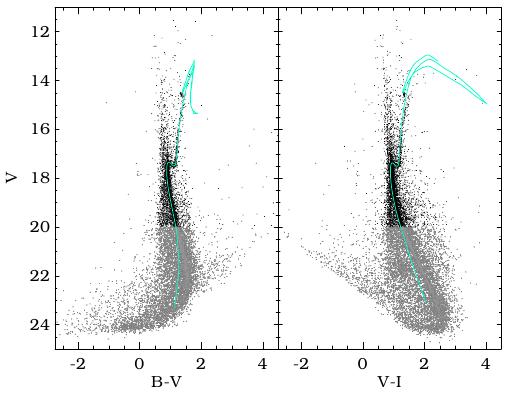}
\caption{CMD of NGC6791 and maximum likelihood isochrone for $V,\ B-V$ (left) and
$V,\ V-I$ (right)}
\label{fig:ngc6791}
\end{figure}

The results for NGC6791 are plotted in Fig.~\ref{fig:ngc6791} and
summarised in Table~\ref{tab:ngc6791}. From Fig.~\ref{fig:ngc6791} the
method appears to have provided a reasonable fit to the data despite
the presence of foreground stars in the CMD.  The likelihood for $BV$
is higher than for $VI$, which is indicative of a better fit to the
data. The $BVI$ results fit the same age and metallicity as the $BV$
results, but have a distance and reddening consistent with the $VI$
results. It should be noted that [Fe/H]~$= +0.4$ is the highest in the
isochrone library, so it is possible that a higher metallicity is more
consistent with the data.  The best set of empirical parameters
(determined from eclipsing binaries) is from \citep{Brogaard11} who
found [Fe/H]~$= +0.29 \pm 0.03$ (random) $\pm 0.07$ (systematic),
$E(B-V) = 0.160 \pm 0.025$ and $\mu_V = 13.51 \pm 0.06$. Our results
are in good agreement with these.  Our results (particularly age) are
also consistent with \citet{Bedin08} who fit the BaSTI isochrones to
HST data; they obtained [Fe/H]~$=$~0.4 and find $E(B-V)=0.17$,
${\mu_V}=13.50$ and $t=8$ Gyr.

\section{Complex Stellar Populations}
\label{sec:test_csp}

Analysing the CMD of a complex stellar population (CSP) poses more of
a challenge than SSPs, since the method not only has to be able to
identify the correct isochrones in the solution but also allocate the
correct number of stars to these isochrones. The synthetic SSP tests
allowed us to investigate the robustness of the method in selecting
the correct isochrone. For the synthetic CSP tests we complete similar
tests, but now the emphasis on whether the method can accurately
reproduce different types of star formation histories. In particular
we want to ascertain how well the method can distinguish between
bursting and continuous star formation scenarios.

\begin{figure}
\centering
\includegraphics[scale=0.7]{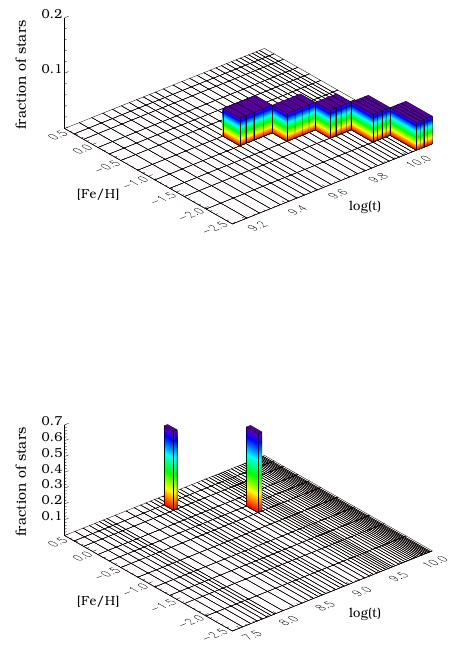}
\caption{Input star formation histories for population 1 (top) and
  population 2 (bottom). The x and y axes represent age and [Fe/H], each
  rectangle representing an isochrone. The vertical axis represents
  the fraction of stars at each age and metallicity, which is
  essentially the weight of each isochrone.}
\label{fig:input}
\end{figure}

We present two star formation histories, depicted in
Fig.~\ref{fig:input}, which will be referred here on in as population
1 and 2. Population 1 represents an old stellar population with
constant star formation rate with ages between 3.5 Gyrs and 13 Gyrs
and metallicities between [Fe/H]~$=-$2.27 and [Fe/H]~$=-$0.96.
Population 2, on the other hand, consists of two equal bursts of star
formation, one with an age of 2 Gyrs at [Fe/H]~$=-$0.35 and the
second with an age of 500 Myrs and [Fe/H]~$=$~0.26.

For both populations 1 and 2 we create a synthetic CMD using BaSTI
synthetic CMD generator \citet{Cordier07}. Simulated photometric
errors are applied to this original CMD using the same error law as
the SSP tests, but with a range of values for $A_{0}$, to test the
method with different quality of photometric data. At this point, we
do not include distance or reddening, but fix $\mu_0=0$ and
$E(B-V)=0$.  We solve for $V$ and $I$, and take a lower MS magnitude
cut at $M_V=6$, which is about 2 magnitudes below the oldest possible
turn-off.  Note that the completeness will be 1 for all magnitudes. We
test the effect of photometric incompleteness in
sec~\ref{sec:imf_compl}.

\begin{figure*}
\centering
\includegraphics[scale=0.6]{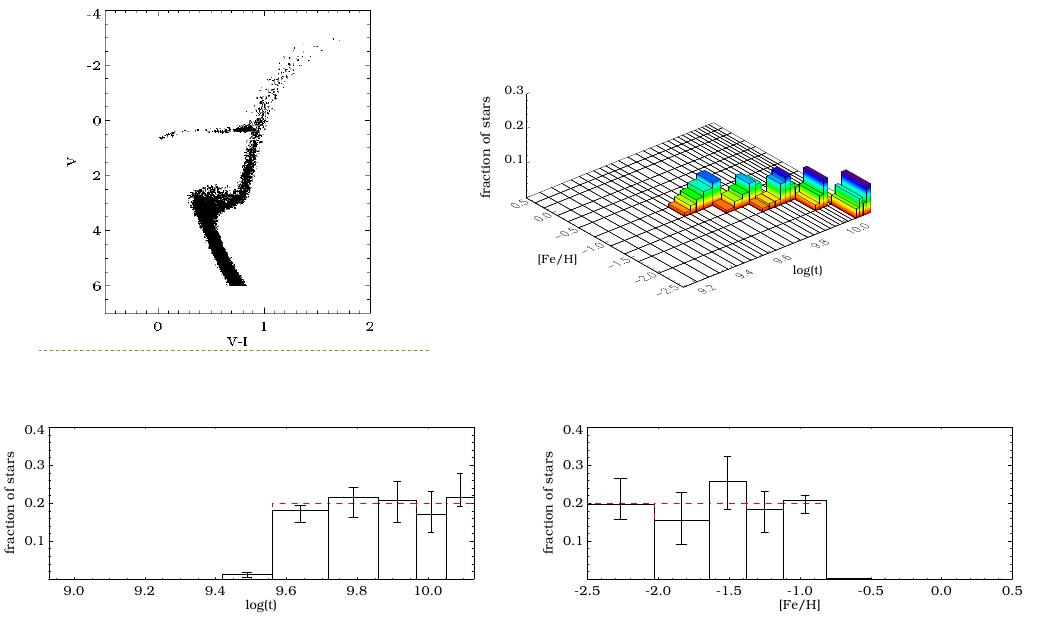}
\caption{Results for pop. 1 ($A_0=5$): \textit{Top left:} input 
  CMD \textit{top right:} plot
  of the full recovered star formation history, \textit{bottom left:}
  plot of the relative star formation rate as a function of time
  \textit{bottom right:} plot of the relative star formation rate as a
  function of metallicity. The input SFH solution is plotted in red
for comparison in the bottom two panels. 
}
\label{fig:pop1}
\end{figure*}

The input CMD and recovered star
formation histories for $A_0=5$ are displayed in
fig.~\ref{fig:pop1} and \ref{fig:pop2} for population 1 and 2 
respectively.

\begin{figure*}
\centering
\includegraphics[scale=0.6]{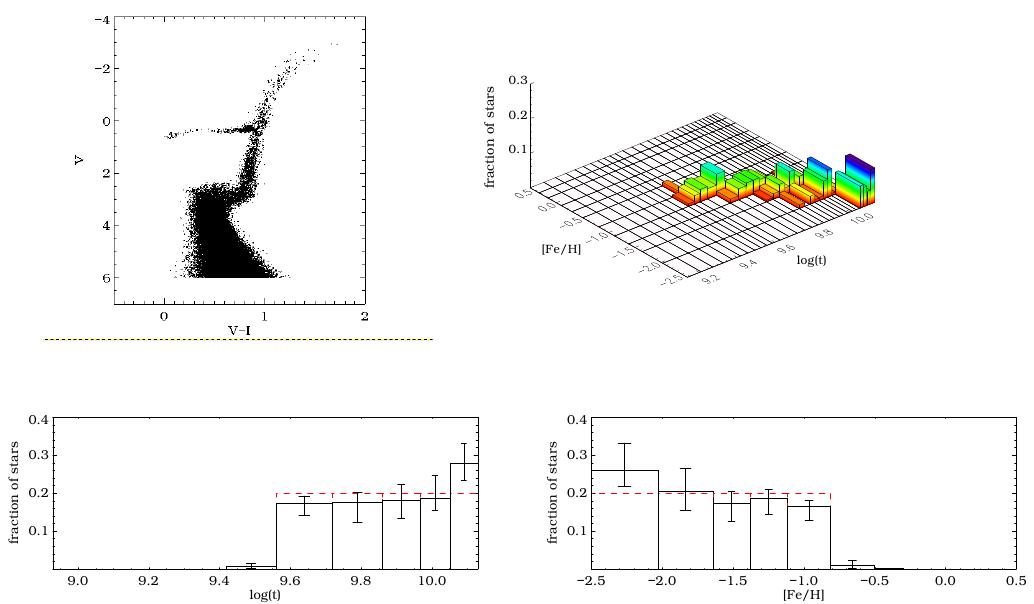}
\caption{Results for pop. 1 with $A_0$~=~-1.5 (same format as in Fig.~\ref{fig:pop1})}
\label{fig:pop1-1.5}
\end{figure*}

\begin{figure*}
\centering
\includegraphics[scale=0.6]{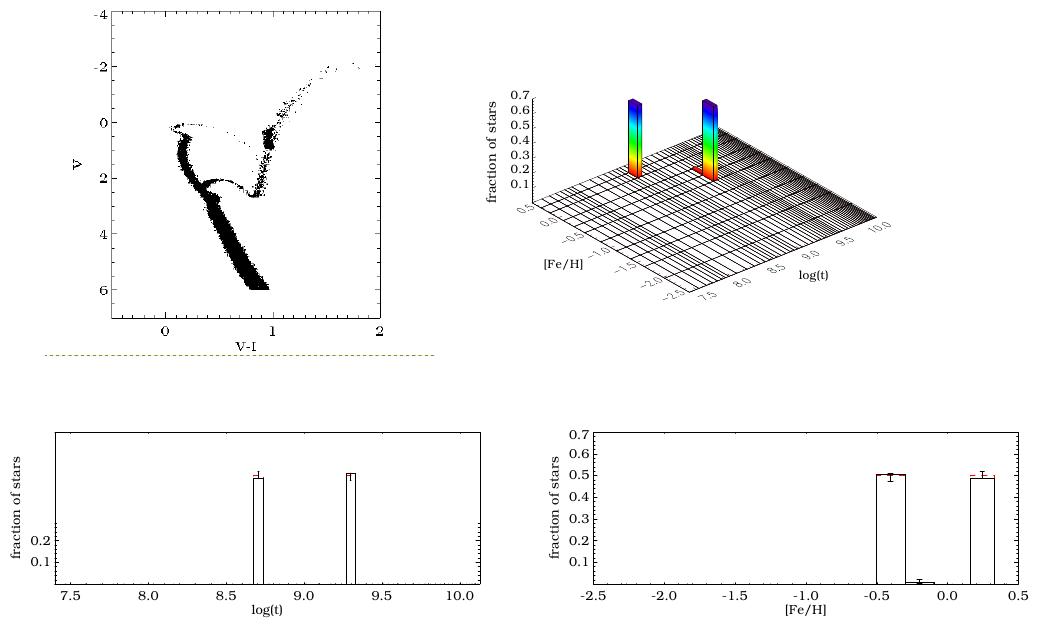}
\caption{Results for pop. 2 (same format as in Fig.~\ref{fig:pop1})
}
\label{fig:pop2}
\end{figure*}

\begin{figure*}
\centering
\includegraphics[scale=0.6]{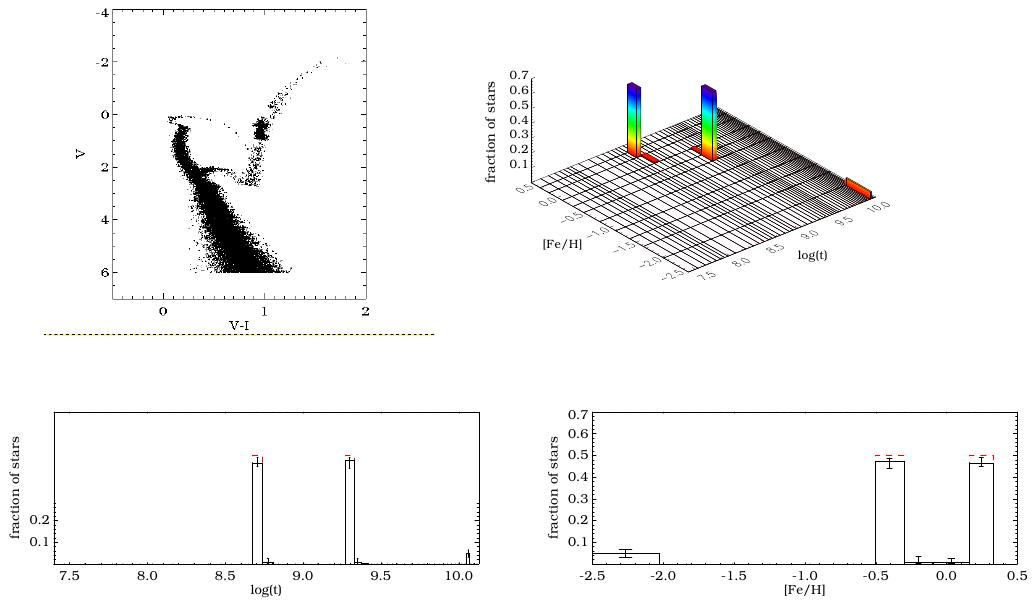}
\caption{Results for pop. 2 with $A_0$~=~-1.5 (same format as in Fig.~\ref{fig:pop1})
}
\label{fig:pop2-1.5}
\end{figure*}

The method was able to reproduce the overall star formation history
for several values of $A_0$ in the error law.  However the weights of
the isochrones fluctuate more from their true values with data that
has a greater photometric error.  This is understandable since the
isochrones that are close in age and metallicity are also close in the
CMD, and large photometric errors diminish the distinction between
these isochrones in terms of their relative probability.  Our tests
for population 1 show that when the photometric errors increase (at a
given magnitude) this causes oscillations or even gaps in the star
formation rate due to stars, originating from one isochrone, being
accounted for entirely by neighbouring isochrones.  This could be
misinterpreted as a varying or bursting star formation rate, so it is
always important to calculate confidence limits and present the
results using suitable age and metallicity bins.

We also test the effect of increasing and decreasing the photometric
errors of stars in the synthetic CMD by factors of 2 and 5 to mimic
the effect of estimating errors inaccurately. Overestimating errors
has the effect of increasing the ``burstiness'' of the star formation
history solution. 

For population
1, although each input metallicity is represented in the solution,
stars become more concentrated on the central age of each
subpopulation. Also more stars are given to the subpopulation with the
mid-range metallicity. For population 2, the bursty star formation is
actually recovered better with the overestimated errors than with the
orginal errors.  On the other hand, underestimating errors increases
the apparent duration of the star formation events and adds spurious
isochrones into the solution.

\begin{figure}
\centering
\includegraphics[scale=0.55]{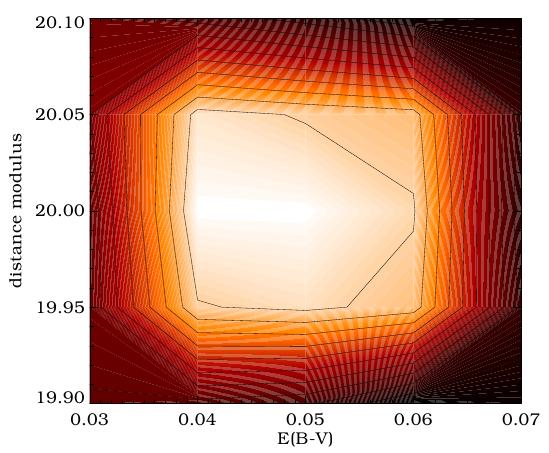}
\caption{Contour plot of likelihood for pop. 1 with input $\mu_0=20,
  E(B-V)=0.05$ against distance and reddening}
\label{fig:likedr}
\end{figure}

We also generate a CMD for population 1 with a distance modulus
$\mu_0=20$ and $E(B-V)=0.05$, to test whether we can use the method to
estimate these parameters as well as determine the star formation
history. We simulate the photometric error using
eq.~\ref{eq:error_law} with $A_0=18.5$. We determine the maximum
likelihood star formation history for distances $\mu_0=\{19.9,19.95,
\ldots,20.1\}$ and reddenings $E(B-V)=\{0.03,0.04,\ldots,0.07\}$.
The maximum likelihood solution recovers the input values of reddening
and distance modulus \textit{as well as} the same SFH as in the
previous case.  Fig.~\ref{fig:likedr} shows how maximum likelihood
changes with distance and reddening values.

\subsection{Unresolved Binaries}

\begin{figure*}
\centering
\includegraphics[scale=0.6]{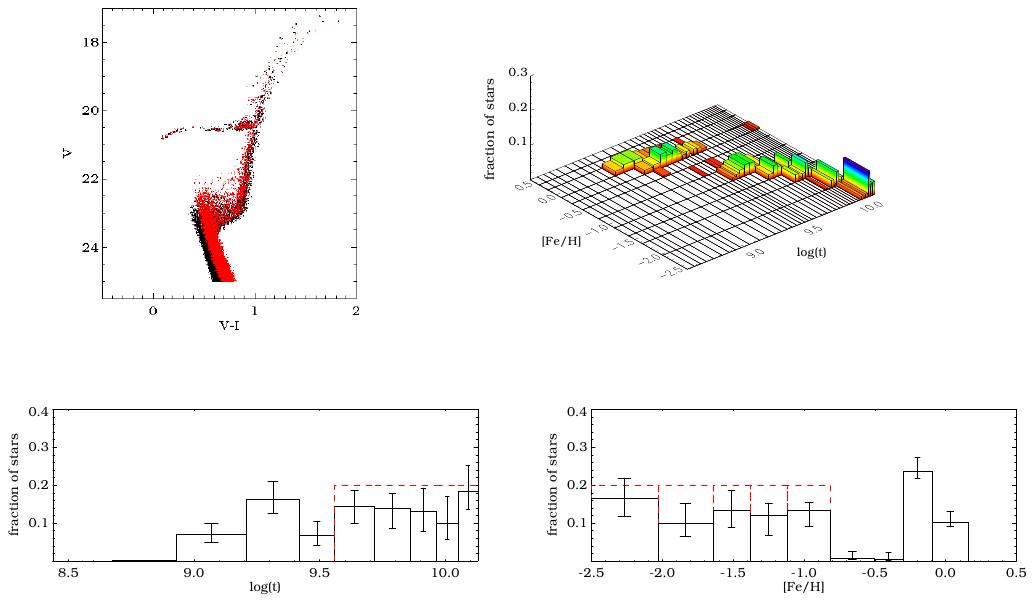}
\caption{CMD and recovered star formation history for pop. 1 with
  binary fraction = 0.4 (same format as Fig.~\ref{fig:pop1}). The red
stars in the CMD are the binaries.} 
\label{fig:pop1_m5_bin}
\end{figure*}

\begin{figure*}
\centering
\includegraphics[scale=0.6]{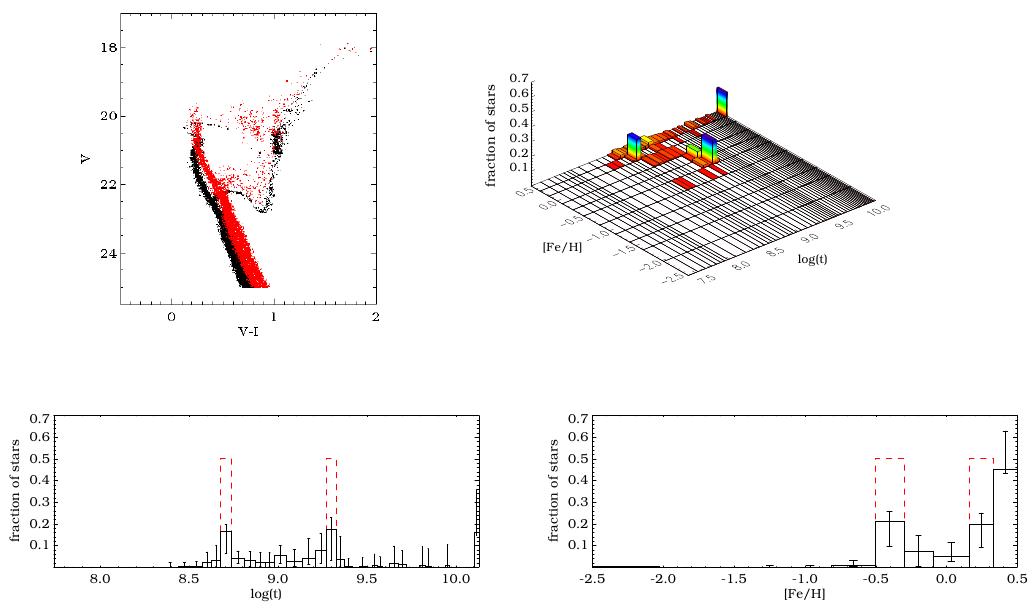}
\caption{CMD and recovered star formation history for pop. 2 with
  binary fraction = 0.4 (same format as
  Fig.~\ref{fig:pop1}). The red stars in the CMD are binaries. }
\label{fig:pop2_m5_bin}
\end{figure*}

The SSP tests showed that the method was sensitive to the effects of
binaries only for a high binary fraction and for young stellar
populations. As we are no longer forcing the method to select a single
isochrone, but to attribute stars to a combination of isochrones, it is
expected that unresolved binaries will have a different effect on the
recovered star formation history. To perform this test, we generate
synthetic CMDs for population 1 and 2 with a binary fraction of 0.4
and the same mass ratio distribution as the SSP tests.  We perform the
test using two error laws: $A_0=5$ and $A_0=-1.5$.  We also apply a
distance modulus $\mu_0=20$ and reddening $E(B-V)=0.05$ to the
synthetic data and simultaneously test how binaries affect these
recovered parameters. The range of values are
$\mu_0=\{19.75,19.8,\ldots,20.05\}$ and
$E(B-V)=\{0.04,0.05,\ldots,0.09\}$.

The CMD and recovered star formation history for population 1 
with $A_0=25$ is displayed in Fig.~\ref{fig:pop1_m5_bin}. 
It is evident from
the CMD that the population of binary stars (red points) on 
the MS are distinct from the single stars (black points). For 
the $A_0=18.5$ case, the photometric error effectively mixes
the two together on the MS. For the correct value for 
$\mu_0$ and $E(B-V)$, the binaries are seen in the maximum 
likelihood solution as an additional population of younger and
more metal rich stars. In Fig.~\ref{fig:pop1_m5_bin}, the
binaries are evidently separate from the rest of the solution,
whereas for $A_0=18.5$ the binaries are seen as an extension of
the main star formation epoch. Nevertheless, the rest of the 
star formation history is recovered quite well in both cases. 
The method estimated
the distance and reddening correctly for $A_0=25$, but there was a
small shift to a lower distance modulus $\mu_0=19.95$ for $A_0=18.5$.
This could possibly be attributed to binaries close to the MSTO. The
shift in $\mu_0$ is too small to have a significant effect on the
recovered star formation history. 

For population 2 we encounter the same problem as we did for the young
SSP case. The fit is dominated by the MS, and the effects of binaries
on the recovered star formation history are more dramatic.
Unfortunately we cannot deal with binaries in the way we did for SSP
since we can only take a magnitude cut below the TO of the oldest
stars in the CMD without losing age information provided by the TO and
SGB.  This still leaves a significant amount of MS binaries in the
data.

The CMD and recovered star formation history of population 2 for
$A_0=18.5$ is presented in Fig.~\ref{fig:pop2_m5_bin}. For the younger
burst, the MS and binary sequence are very distinctly separated in the
CMD for both error laws. If $\mu_0$ and $E(B-V)$ are fixed to the
correct values, the centres of two bursts are recovered correctly in
the maximum likelihood solution.  However stars are also attributed to
other isochrones, particularly those with the highest metallicity. The
age spread at this metallicity may be due to the fact that the younger
burst is close to the edge of our ischrone grid, so that binaries too
red to be attributed to a single MS are being fit by MSTO and subgiant
branch of older isochrones instead.

If the distance modulus and reddening are left as free parameters then
the maximum likelihood solution is at $\mu_0=20$ and $E(B-V)=0.05$ for
$A_0=25$ and $\mu_0=19.75$ and $E(B-V)=0.07$ for $A_0=18.5$.  We note
that $\mu_0=19.75$ is the lowest in our range and the actual maximum
solution may be at an even lower distance. If this is the case then it
suggests that distance and reddening parameters are not being
constrained at all by the post-MS stars.  The derived star formation
history is also affected by the incorrectly estimated parameters.  The
centres of the two bursts are shifted to a lower metallicity
[Fe/H]~$=$~0.06 and [Fe/H]~$=-$0.66
to compensate for the higher reddening.  The age of the younger burst
is also lower, although the age of the older burst remains the
same. The reason that distance and reddening results are only affected
in the $A_0=18.5$ case is that the photometric error effectively blurs
the distinction between single and binary stars and gives greater
leeway in fitting the isochrones.  The parameters are skewed as the
method attempts to find an average fit for both populations. On the
other hand, single and binary stars remain distinct when the
photometric errors are small and the overall distance and reddening
parameters of the whole population are dominated by the more numerous
single stars.

To summarise, 
binaries do affect the recovered star formation history by taking
stars away from correct isochrones and populating younger, more metal
rich isochrones.  Binaries also transform the likelihood landscape for
distance and reddening, causing it to become more complex and
multi-peaked compared to Fig.~\ref{fig:likedr}. 
We have completed the
tests using a relatively high binary fraction and expect similar but
less significant effects for smaller binary fractions.  The effects of
unresolved binaries cannot be ignored when using the method and must
be a caveat in interpreting the star formation history of a stellar
population with an unknown binary fraction.  We will leave the subject
of dealing with unresolved binaries in a fully consistent way as the
subject of a future paper.

\subsection{HB Morphology}
\label{sec:csp_hb}

Similar to our SSP test, we explore the effect of an incorrect HB
morphology, by creating a CMD using $\eta =0.4$ isochrones and solved
using $\eta=0.2$.  In the SSP tests we found that a bluer HB could
lead to a more metal poor and subsequently older isochrone being
selected. The point of repeating these tests for complex stellar
populations is to ascertain the effect of blue HB stars when we remove
the constraint that all stars of a population belong to one
isochrone. We need to establish how important HB stars are in
determining the star formation history. We also want to check whether
blue HB stars can possibly be mistaken for young MS stars, since there
can be an overlap of isochrones in the CMD.

Again we complete the tests for population 1 using the two cases of
simulated photometric errors, $A_0=5$ and $A_0=-1.5$. We actually take
a magnitude cut of $V=4$ rather than $V=6$ to increase the weight of
the HB in star counts in the CMD. We evaluate the star formation
history with and without the inclusion of the HB stars. There is no
apparent difference between the two results for $A_0=5$, leading us to
the conclusion that the HB is effectively ignored in this
instance. For $A_0=-1.5$ we find that the star formation history is
recovered slightly better when the HB is removed. The actual effect is
a slight metallicity spread at some of the ages and is relatively
small compared to fluctuations due to photometric error.  For neither
case do we falsely detect young stars due to overlap of HB stars with
young isochrones.

These tests reveal that the star formation can be determined reliably
without fitting HB stars. They also show that a different HB
morphology can only bias the results if there is greater leeway in the
fit to the MS by photometric error. This is further confirmation that
the method is very much dominated by the lower mass stars in the CMD.

\subsection{IMF and Completeness}
\label{sec:imf_compl}

\begin{figure}
\includegraphics[scale=0.5]{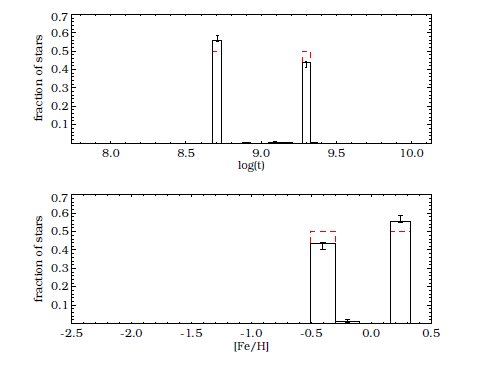}
\caption{Plot of the relative star formation as a function of time
   and metallicity for pop. 2 with simulated incompleteness (see text for details).}
\label{fig:pop2_comp}
\end{figure}

Our SSP tests showed that the IMF and completeness do not affect the
selection of the correct isochrones.  However, since IMF and
completeness determine the distribution of stars along each isochrone in
the CMD, it is important to check that these factors do not
significantly affect the isochrone weights in the maximum likelihood
solution.  Especially since we need to correct for our assumed IMF and
completeness to get the relative number of stars formed at each age
and metallicity.

For the IMF test, we populate isochrones in a synthetic CMD of our
population 2 using an IMF with a power law exponent $-2.35$. 
In these tests, we simulate photometric errors using $A_0=5$ so as to
minimise the fluctuations in the weights caused by the photometric
errors.  The star formation history is then evaluated assuming
exponents of $-1.35$, $-2.35$ and $-3.35$.  We note that changing the slope of
the IMF power law will alter the relative probabilities of the
individual stars.  Increasing the exponent effectively increases the
probabilities of lower mass stars compared to higher mass stars,
thereby increasing their contribution to the likelihood function, and
vice versa. However, our tests show that the unscaled weights for each
isochrone are not affected by different IMF exponents, although the
maximum likelihood is greatest for the correct value. We infer that
the method is robust against an incorrect IMF.
 
Rescaling with an
incorrect IMF will affect the final isochrone weights. 
We will underestimate (respectively overestimate) the number of stars
needed to correct for the older burst if the IMF slope is too steep
(resp. too shallow).
Although this effect will vary depending on the specifics of
the star formation history solution, we find that for our synthetic
test the change in the weights are of the order of $5-10\%$.

We now introduce the effect of incompleteness, using
equation~\ref{eq:comp} with $A_c=5$ and $\Delta A=0.45$ to remove
stars from our previous CMDs for population 2. Again we simulate the
errors only using $A_0=5$.  The results show that the weight of the
older burst is underestimated and the younger burst is overestimated
by $\sim 5-10\%$ (see Fig.~\ref{fig:pop2_comp}).  
At a given magnitude (below the TO of the oldest population) the more
metal-poor MS is populated by lower mass stars than the metal-rich
component.  Due to the shape of the IMF, we are therefore missing
more metal-poor stars than metal-rich stars.
If completeness is accounted
for then rescaling to a given mass range will correct for this
imbalance.  One way of avoiding the problem of incompleteness is to
take a cut on the MS, above which one can be sure that the data will
be complete or at least has no magnitude dependence due to measurement
limitations.

\section{Carina Dwarf Galaxy}
\label{sec:carina}

The final test is to apply the method to a CMD of a real complex
stellar population, the Carina dwarf galaxy.  There have been many
studies of its stellar populations in the literature, both
spectroscopic and photometric, with which we can compare our results
and there is general agreement regarding its star formation history.
Our aim here is not to perform a detailed analysis of the Carina CMD
to recover the SFH but to test whether we can recover what is fairly
well known about this galaxy.

Studies of the SFH of Carina have identified several episodes of star
formation: an initial burst at $\sim 12$~Gyr followed by an
intermediate-age event at $\sim 6$~Gyrs, with a possible more recent
episode at $\sim 2$~Gyr \citep{Smeckerhane96,Hurley98,Rizzi03}. The
intermediate age episode however can have a different duration
according to different authors and this may affect the existence of
the most recent one (as it ``merges'' with the intermediate age
population).  Studies based on HST data have found a more continuous
SFH with one or more ``peaks'' \citep{Mighell97,Hernandez00,
Dolphin02} but they
do not detect a significant old population (because of the small
number of stars in the field).  Overall it is clear however that there
was an ancient episode of star formation followed by one and possibly
two others.

Because of its proximity, it is also possible to obtain spectroscopy
of red giant stars in Carina to study its metal enrichment history.
The consensus is that Carina is a fairly metal-poor system with an
average [Fe/H] around $-1.6$ \citep{Koch06,Koch08,Tolstoy03,Lemasle12}.  The
may be a range a range in metallicity of order 0.5 dex
\citep{Tolstoy03,Lemasle12}  or even higher depending on the method
used \citep{Koch06}.  Photometric studies tend to agree on the fact
that the metallicity range has to be fairly small
\citet{Smeckerhane96,Rizzi03,Bono10} in order to reproduce the
thinness of the RGB. More recently, Fabrizio et al. (2012) 
measure a more modest spread and a
more metal poor mean [Fe/H] from spectroscopy of red giant 
stars, which is in better agreement with photometric studies.

\begin{figure}
\includegraphics[scale=0.7]{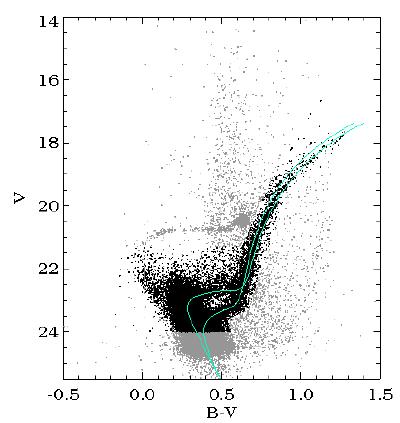}
\caption{CMD for the \citet{Bono10} dataset. The grey points
  represent stars removed from the dataset prior to the SFH
  determination}
\label{fig:car_cmd}
\end{figure}

We use the data set of \citet{Bono10} which covers a fairly wide field
and reaches well below the MS turn off of the oldest population.  The
CMD is already background subtracted with an algorithm described in
that paper.  The $V,\ B-V$ CMD is shown in Fig.~\ref{fig:car_cmd},
where black points are included and grey points have been removed from
the dataset prior to analysis. 
Carina exhibits an extended blue HB as well as a red clump but, to
avoid problems with uncertain predictions of HB morphology, we are
neglecting the helium-burning phase in our analysis.
We have also taken a lower MS
magnitude cut at $m_V=24$. We have not determined the completeness

The estimated photometric errors for this dataset of Carina are small
enough that the total error is dominated by $\sigma_{\mathrm{iso}}$.
We solve for distance modulus and reddening simultaneously with
initial values of $\mu_0 = \{19.9,19.95,\ldots,
  20.2\}$ and $E(B-V)=\{0.03,0.04,\ldots,0.08\}$. We have used
the solar-scaled isochrones since spectroscopic studies have shown
evidence of very little $\alpha$-enhancement in the majority of
Carina's giant stars and other dwarf spheroidal galaxies (e.g. Shetrone et al.
 2003). We have solved for $BV$, $VI$ and $BVI$, however we
present the $BV$ results since our likelihood calculations indicate a
better fit to the data than $VI$ or $BVI$.
We have attempted to solve for the star formation history while
leaving distance modulus and reddening as free parameters. 
For $BV$ our maximum likelihood solution is at $\mu_0=20$ and
$E(B-V)=0.06$.

\begin{figure}
\centering
\includegraphics[scale=0.6]{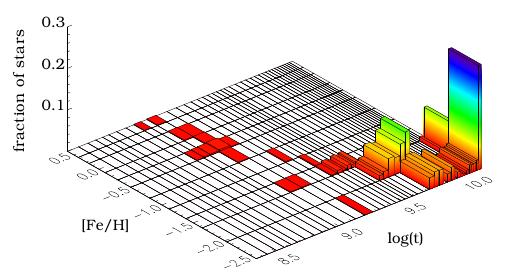}
\caption{The star formation history of Carina using $B,\ V$ (isochrone weights have been rescaled). The cluster of metal-rich isochrones near 2~Gyr is likely due to binary stars (see text for details).}
\label{fig:car_sfhbar}
\end{figure}

\begin{figure}
\includegraphics[scale=0.55]{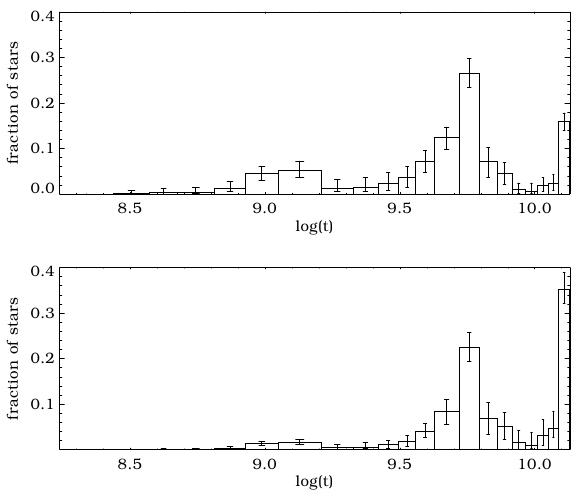}
\caption{The star formation rate as a function of age for Carina based
  on our $BV$ results.  The top plot shows the maximum likelihood
  solution. The bottom plot shows the same but after having rescaled
the isochrones to determine the relative number of stars  
formed at each age.}
\label{fig:car_sfr}
\end{figure}

\begin{figure}
\includegraphics[scale=0.55]{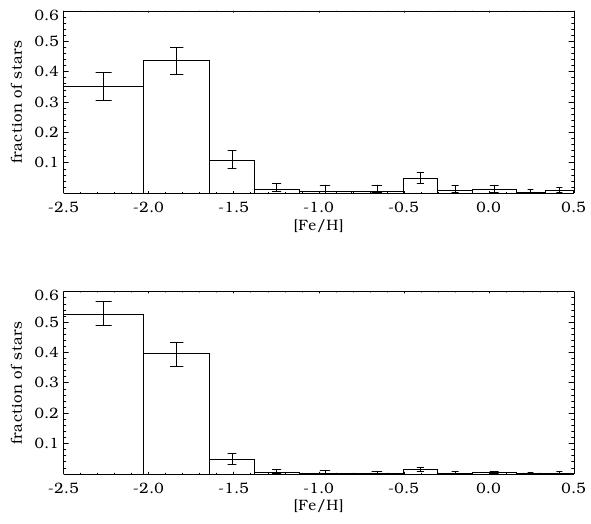}
\caption{Metallicity distribution for Carina based on our $BV$
  results. The top plot shows the maximum likelihood solution. The
  bottom plot shows the same but after having rescaled the isochrones.}
\label{fig:car_ce}
\end{figure}

We present the star formation history in full in
Fig.~\ref{fig:car_sfhbar} and as a relative star formation rate as a
function of age and metallicity in Figs.~\ref{fig:car_sfr} and
\ref{fig:car_ce}. Our star formation history is dominated by two
bursts; one centred on (13~Gyrs, -2.27) and one centred on (6~Gyrs,
-1.79). We overplot the central isochrones representing each burst on
the CMD of Carina in Fig.~\ref{fig:car_cmd}. Here we are using the
term burst loosely to describe enhanced periods of star formation,
lasting several Gyrs, interspersed with a period of negligible star
formation.

Our results qualitatively agree with those of \citet{Smeckerhane96}
and \citet{Hurley98} and most closely resembles that of
\citet{Rizzi03}. We find that the second burst formed most of its
stars at $\sim 6$~Gyrs but there is an extended tail of star formation
until $\sim 3$~Gyrs.

According to our results, $34\%$ of stars in the CMD were formed at 13
Gyrs while $46\%$ were formed between 4.5 and 7~Gyrs.    
Our solution also contains a small
number of more metal rich lower mass stars, around 1 Gyrs. However if
we overplot the isochrone (1 Gyr,$=-$0.35) it is evident that there
are no RGB or He burning stars belonging to this population evident in
the CMD. We suspect that these are actually unresolved binaries from
the more metal poor star formation events rather than a genuine metal
rich sub population. If this is the case then the number of stars
associated with our intermediate burst are slightly underestimated.

One of the benefits of our method is that the chemical enrichment
history does not have to be assumed but can be determined
self-consistently with the star formation rate.  The results indicate
that Carina consists of mainly metal poor stars, with an overall mean
metallicity lying between [Fe/H]~$=-$2.27 and [Fe/H]~$=-$1.79,
although weighted towards the latter.  This is consistent with several
studies cited above.  It is lower but still compatible with
\citet{Bono10}, who measure a metallicity of [Fe/H]~$\simeq -1.70\pm
0.19$ dex, from the difference in colour between the red clump and the
middle of the RR Lyrae instability strip.

Although in our results both populations have stars at both
metallicities, more old stars are at [Fe/H]~$=-$2.27 and more
intermediate age stars are at [Fe/H]~$=-$1.79, in agreement with
\citet{Lemasle12}.  This indicates that some chemical enrichment
occured between the two bursts, although it is hard to quantify with
the metallicity resolution of the isochrone grid. Our results also
show a spread in metallicity within each burst; the intermediate age
burst in particular has stars associated with [Fe/H]~$\sim -2.27$ to
[Fe/H]~$\sim -1.49$. Although, again we have to bear in mind our
metallicity resolution, which could make this spread appear larger
than it actually is.  Our recovered star formation history shows an
age metallicity dependence during the second period of star formation
in that more metal rich isochrones are also younger. Also, since our
result is also constrained by the MS and SGB, the age metallicity
dependence seen is not just due to age metallicity degeneracy on the
RGB.

\section{Summary and conclusions}
\label{sec:conlusions}

We have presented a method to determine the star formation history
(including its metallicity history) of any stellar population. We
model the star formation history as a linear combination of simple
stellar populations and determine the relative number of stars
belonging to each sub population using a maximum likelihood
method. Our likelihood function is constructed from the combined
probabilities of all the stars in the CMD originating from the
isochrones in our grid, fully taking into account individual
measurement errors, IMF and completeness.  The maximum likelihood is
determined using all evolutionary phases in the CMD, although the fit
is naturally weighted towards fitting the lower mass stars. However
there is the option of removing problematic features from the CMD,
such as the lower MS and HB; the star formation history is recovered
reliably without these stars.  The method can be used with fixed or
free distance and reddening parameters. Distance and reddening can be
solved for simultaneously by determining the star formation history of
each combination of parameters and taking the highest of the maximum
likelihoods.  If the method is being used to estimate distance and
reddening, it is recommended to check that the isochrones in the
solution are consistent with the data when overplotted on the CMD.

The method has several advantages: \textit{i)} it does not rely on any
assumption regarding the star formation history or age-metallicity
relation \citep{Hernandez99}, \textit{ii)} it does not require the
calculation of elemental CMDs \citep{Aparicio09,Harris01,Dolphin02,deBoer12}
but compares the data directly to the isochrones, \textit{iii)} it
does not parameterise the CMD by binning, but treats each star
individually. This last point means that the method has the potential
to recover very detailed star formation histories, since no
information is being lost through the binning process.  The maximum
possible resolution in time and metallicity of the solution will
depend on the isochrone grid, however, in practice it will usually be
set by the photometric errors. Also, unlike parameteric methods, the
method doesn't necessarily require an IMF to be fully sampled by the
population, since the probabilities are taken as an integral over the
whole isochrone. Therefore, the method will not be as sensitive to
small number statistics and will be more reliable for datasets
containing fewer stars.

We have tested the method extensively using synthetic and real data
for both simple and complex stellar populations.  We have shown that
the method is relatively robust against most systematics, including
background contamination, HB morphology, IMF and
completeness. Unresolved binaries are the only systematic effect that
may have a significant effect on the recovered star formation
history. Unlike the synethetic CMD method, which can include a binary
fraction in the computation of elemental CMDs, isochrones only model
single stars. In the star formation history solution, binaries are
seen as an additional young metal rich populations, which in many
cases, such as Carina, are separate from the main star formation
events. A fully consistent way of treating unresolved binaries is
beyond the scope of this paper, but it may be possible to correct for
binaries if they can be isolated and identified in SFH solution. 
Since binaries will only affect the MS, in certain
situations one indicator that the isochrone is not genuinely part of
the solution will be the absence of post-MS stars in the CMD.

One of the most promising aspects of the method is that it is easily
extended to datasets where there are more than two
magnitude measurements per star. The probability function can be
adapted to include multiple magnitudes, allowing all the information
available for each star in the CMD to be used to constrain the star
formation history.  Another possibility is to derive the star
formation history of the data using different colour combinations.
Analysis of the separate probabilities for the stars can then
highlight inconsistencies between the isochrones in different
passbands and can possibly be used to test stellar evolution models.

\section*{Acknowledgments}

We thank J. Hartman for the M~37 data.  We acknowledge the Carina
Project for sending us the optical catalog in electronic form.  We
thank the referee for a report that improved the content and
presentation of this paper.  EES was supported by a studentship grant
from STFC.

\end{document}